\PassOptionsToPackage{unicode}{hyperref}

\documentclass[cameraready]{Interspeech}
\usepackage[T1]{fontenc}
\usepackage{url}
\usepackage{subcaption}

\usepackage{booktabs}
\DeclareRobustCommand{\eth}{\dh}
\AtBeginDocument{%
  \pdfstringdefDisableCommands{%
    \def\eth{ð}%
  }%
}

\title{Improving multichannel speech enhancement through accurate room-acoustic simulations}

\author[affiliation={1}, orcid=0000-0003-2896-5341, correspondingauthor]{Georg}{Götz}
\author[affiliation={1}, orcid=0000-0003-1862-5087]{Alessia}{Milo}
\author[affiliation={1}, orcid=0009-0008-0974-7705]{Steinar}{Gu\eth{}jónsson}
\author[affiliation={1}, orcid=0000-0002-7762-6568]{Daniel Gert}{Nielsen\,}
\author[affiliation={1}, orcid=0009-0003-9794-9123]{Jesper}{Pedersen}
\author[affiliation={1}, orcid=0000-0003-2779-2901]{Finnur}{Pind}

\address{
    $^1$ Treble Technologies, Reykjavík, Iceland
}

\email{georg.goetz@treble.tech, am@treble.tech, sg@treble.tech, dgn@treble.tech, jp@treble.tech, fp@treble.tech}

\keywords{Multichannel speech enhancement, far-field speech recognition, room acoustics, geometrical acoustics, wave-based simulation, data augmentation}

\usepackage{comment}

\begin{document}
\maketitle

\begin{abstract}
Room-acoustic simulations are widely used to augment training data for deep-learning-based speech enhancement. While most pipelines rely on simplified geometrical acoustics, wave-based approaches offer greater physical accuracy. In this work, we examine how simulation fidelity affects multichannel speech enhancement performance. To this end, we train SpatialNet on datasets augmented with different room-acoustic simulation methods and evaluate the resulting models on measured data. We compare lower-fidelity datasets based on geometrical acoustics with a high-fidelity dataset using advanced acoustic modelling and a hybrid combination of wave-based and geometrical acoustics simulations. Training on the high-fidelity dataset results in an up to \SI{38}{\percent} relative reduction in median word error rate compared to the lower-fidelity alternatives. These results show that augmentation with high-fidelity room-acoustic simulations directly translates into improved multichannel speech enhancement performance.
\end{abstract}

\section{Introduction}

As smart devices become deeply embedded in everyday life, natural language user interfaces (NLUIs) have emerged as a key paradigm for enabling seamless and intuitive human-device interaction. However, this interaction frequently occurs in acoustically complex environments. Across homes, public spaces, and in-vehicle settings, the intelligibility of user instructions can be significantly degraded by background noise, reverberation, or overlapping speech. To maintain reliable performance under such conditions, automatic speech recognition (ASR) systems rely on speech enhancement techniques like denoising and dereverberation to isolate and preserve linguistic information.

Recent advances in speech enhancement have been largely driven by deep-learning-based methods. Early neural architectures relied on convolutional \cite{Park2017CNNforSpeechEnhancement} and recurrent networks \cite{Weninger2015SpeechEnhancementWithLSTM} to learn time-frequency masks or directly estimate clean speech spectra \cite{Wang2018SupervisedSpeechSeparationDeepLearning}. More recent models incorporate attention mechanisms and Transformer-based architectures to capture long-range temporal dependencies and complex acoustic patterns. For instance, deep filtering approaches like DeepFilterNet \cite{schroter2022deepfilternet} combine deep neural networks with classical signal processing principles for real-time enhancement. In contrast, hybrid architectures like SpatialNet integrate convolutional layers with multi-head self-attention mechanisms and leverage both local spectral structure and global contextual modelling \cite{Quan2024SpatialNet}. These data-driven techniques have substantially improved robustness in challenging acoustic conditions and now represent an important paradigm in contemporary speech enhancement research.

Going beyond the single channel paradigm, multichannel speech enhancement exploits additional information captured by microphone arrays. This additional information enables systems to leverage spatial diversity between target speech and interfering sources to improve robustness in noisy environments~\cite{haebumbach2024microphonearrays,zeng2024timefrequency}. By jointly exploiting spatial and spectral characteristics of the recorded signals, multichannel approaches can achieve improved separation and enhancement performance compared to single-channel methods \cite{haebumbach2024microphonearrays,zeng2024timefrequency}.

Deep learning models are trained on increasingly large and heterogeneous datasets, such as those introduced in the Deep Noise Suppression (DNS) challenges \cite{dubey2024icassp}. In practice, data augmentation is typically performed by convolving clean speech with room impulse responses (RIRs) to simulate different acoustic environments. However, much of the state-of-the-art still relies on relatively simplistic synthetic RIRs, which are often derived from shoebox geometries rendered with the image-source method (ISM) using a single frequency-independent absorption coefficient \cite{Ko2017DataAugmentationReverberantSpeechForSpeechRecognition, Kim2017GenerationLargeScaleSimulatedUtterancesFarFieldASR,Maciejewski2020WHAMR}. However, such simple RIRs may not sufficiently capture the acoustic complexity of real environments. 

Prior evidence in related tasks suggests that increasing RIR realism can improve downstream performance \cite{bezzam2020study, arakawa2024quantifying, Srivastava2023HowToVirtuallyTrainYourSpeakerLocalizer, guso2025mb, Tang2022GWALargeDatasetForAudioProcessing}. Most previous studies focused on the effect of source and receiver realism \cite{arakawa2024quantifying,Srivastava2023HowToVirtuallyTrainYourSpeakerLocalizer,guso2025mb} and wall material realism~\cite{bezzam2020study,arakawa2024quantifying,Srivastava2023HowToVirtuallyTrainYourSpeakerLocalizer, guso2025mb}. Additionally, Bezzam et al. investigate the performance difference between ISM- and ray-tracing-based training datasets \cite{bezzam2020study}. Tang et al. also include single-channel numerical simulations in their study \cite{Tang2022GWALargeDatasetForAudioProcessing}. However, none of these studies have explored the effect of simulation fidelity for multichannel speech enhancement with rigid arrays. 

This study aims to close this gap by examining how room-acoustic simulation fidelity affects multichannel speech enhancement performance in real-world conditions. We train \mbox{SpatialNet \cite{Quan2024SpatialNet}} with several training datasets of varying fidelity and evaluate the resulting models on measured data. The remainder of this paper is structured as follows. Section~\ref{sec:SimulationParadigms} summarizes different room-acoustic simulation paradigms. Section~\ref{sec:experiment} describes the experimental setup of this study. Section~\ref{sec:results} presents the results and Section~\ref{sec:conclusions} concludes the paper.

\section{Simulation paradigms for acoustic modelling}
\label{sec:SimulationParadigms}
In the context of data augmentation for speech enhancement, different simulation methods/fidelity may lead to different spatial and spectral characteristics in the generated RIRs. Room-acoustic simulation methods are commonly divided into geometrical acoustics (GA) and wave-based approaches, with hybrid methods combining elements of both. GA techniques, including the ISM and ray-based methods such as ray tracing, approximate sound propagation using specular reflections and energy transport assumptions that become increasingly valid at higher frequencies \cite{Savioja2015OverviewGA,Krokstad1968CalculatingTheAcousticalRoomResponseByTheUseOfRayTracing, Krokstad1983Fifteenyearsexperiencewithcomputerizedraytracing,Allen1979ImageMethodForEfficientlySimulatingSmallRoomAcoustics,Borish1984ExtensionOfTheImageModelToArbitraryPolyhedra}. However, GA does not inherently model wave phenomena such as room modes, or diffraction around edges and obstacles, and diffraction in particular remains difficult to represent in GA-based frameworks \cite{Torres2001ComputationEdgeDiffractionRoomAcousticSimulations,Schissler2014HighOrderDiffractionAndDiffuseReflectionsForInteractiveSoundPropagation}. These limitations are also reflected in discussions of simulation uncertainty and cross-tool variability~\cite{Vorlaender2013ComputerSimulationsRoomAcoustics,Brinkmann2019RoundRobinOnRoomAcousticalSimulationAndAuralization}. 

In contrast, wave-based methods (e.g., finite-difference time-domain, finite-/spectral-element, and discontinuous Galerkin formulations) solve the acoustic wave equation directly, enabling the representation of diffraction, modal behaviour, and complex, frequency-dependent boundary conditions. These effects are particularly relevant at low and mid frequencies where wavelengths are comparable to room dimensions \cite{Hamilton2016FiniteDifferenceVolumeWaveBasedRoomAcousticsThesis,Pind2019SpectralElementRoomAcousticSimulation,Prinn2023ReviewFEMRoomAcoustics,BottelDooren1995FDTDSimulationRoomAcoustic}. While traditionally associated with higher computational cost, recent advances in time-domain formulations and high-performance implementations have made large-scale wave-based or partially wave-based simulations more feasible \cite{Wang2019DiscontinuousGalerkinRoomAcoustics,Pind2020TimeDomainSimExtendedReactingPorousDGFEM,Melander2024MassivelyParallelGalerkinRoomAcoustics}. 

Hybrid approaches combine these paradigms by applying a wave solver below a crossover frequency and GA above it, thereby capturing low-/mid-frequency wave effects while relying on GA assumptions at higher frequencies. The trade-offs between “rays or waves” have been discussed extensively in the room-acoustics literature \cite{Siltanen2010RaysOrWavesStrenghtsAndWeaknesses}. In the context of data augmentation for multichannel speech enhancement, hybrid simulation is intended to reduce discrepancies related to low-frequency modal behaviour, diffraction, and frequency-dependent decay, while maintaining broadband coverage.

\section{Experiment setup}
\label{sec:experiment}
\subsection{Network configuration}
\label{subsec:network_config}
We train SpatialNet \cite{Quan2024SpatialNet} to investigate the relationship between room-acoustic simulation accuracy and neural network downstream performance. Inspired by the Conformer architecture~\cite{Gulati2020Conformer}, SpatialNet integrates convolutional modelling and multi-head self-attention \cite{Vaswani2017AttentionIsAllYouNeed} for end-to-end multichannel speech enhancement in the STFT domain. It combines narrow-band blocks for speaker clustering and temporal filtering with cross-band blocks for learning correlations across frequencies~\cite{Quan2024SpatialNet}. Our experiment uses the SpatialNet-small configuration at \SI{16}{\kilo\Hz} sampling rate. Direct-path speech is used as the target for all training configurations.

SpatialNet is microphone-array-dependent, i.e., speech enhancement on different arrays requires retraining the network for each array geometry \cite{Quan2024SpatialNet}. In this work, we focus on a six-channel subset of the em32 Eigenmike array. More precisely, we choose the channels 1, 19, 11, 27, 21, and 9, as defined in the manufacturer's data sheet\footnote{\url{https://eigenmike.com/sites/default/files/documentation-2023-10/EigenStudio\%20User\%20Manual\%20R02D.pdf}}, corresponding approximately to locations in the front, back, right, left, top, and bottom of the array, respectively. 

Although the Eigenmike is not a typical array for speech enhancement, we specifically chose it for two reasons. First, we want to investigate whether accurate receiver modelling improves speech enhancement compared to simplified open array simulations. Speech enhancement is commonly used in smart speakers consisting of several microphones mounted on a rigid scatterer, and the Eigenmike is a reasonable approximation of this setup. Second, we want to investigate the simulation-to-real performance achievable when training datasets are generated with modern simulation tools. Practical speech enhancement algorithms must operate well under real-world conditions, and evaluation datasets should be designed to assess such scenarios. Several datasets containing Eigenmike measurements are publicly available, thus making the array suitable for reproducible studies of real-world performance. 

\subsection{Training datasets}
\begin{figure}[tbh!]
  \centering
  
  \begin{subfigure}{\columnwidth}
      \centerline{\includegraphics[width=0.8\columnwidth, trim={0 2em 0 0em}, clip]{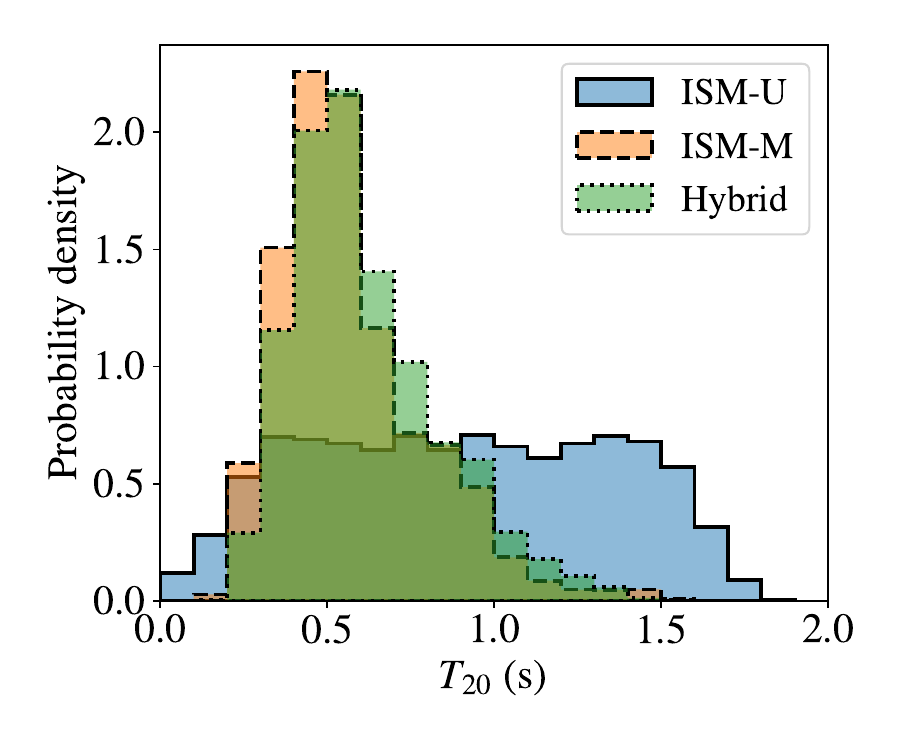}}
      \caption{Training datasets ($N_\mathrm{Scenes} = 4801$)}
      \label{subfig:t20_train}
  \end{subfigure}
  
  \vspace{1em}
  
  \begin{subfigure}{\columnwidth}
      \centerline{\includegraphics[width=0.8\columnwidth, trim={0 2em 0 0em}, clip]{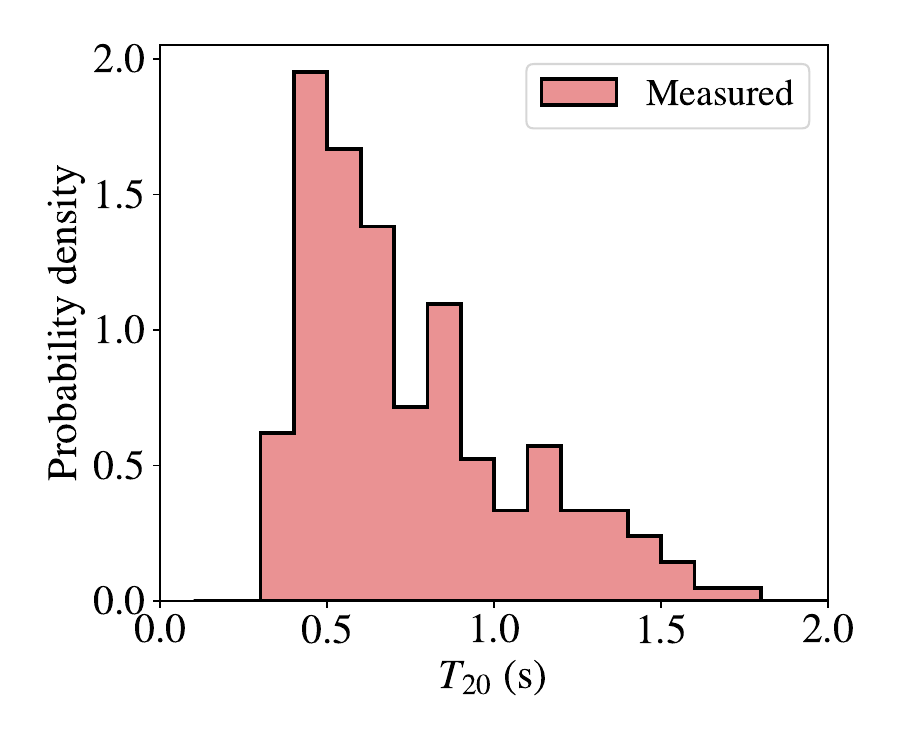}}
      \caption{Evaluation dataset ($N_\mathrm{Scenes} = 60$)}
      \label{subfig:t20_eval}
  \end{subfigure}
  
  \caption{Distribution of the reverberation time $T_{20}$ for the training and evaluation datasets used in this study, averaged over octave bands from \SI{63}{\Hz} to \SI{4}{\kilo\Hz}.}
  \label{fig:t20}
\end{figure}

We train SpatialNet with several datasets augmented using different room-acoustic simulation methods. To this end, we compare lower-fidelity datasets based on GA with a high-fidelity dataset employing hybrid simulations. For the GA-based datasets, we further analyse the effect of dataset design by contrasting an uninformed dataset, where target reverberation times and room dimensions are randomly sampled from predefined ranges, with an informed dataset, where these parameters are matched to realistic material properties and room configurations drawn from curated libraries. Figure \ref{subfig:t20_train} shows the resulting distributions of the reverberation time $T_{20}$ for all training datasets.

\subsubsection{ISM: Image-source dataset}
The original SpatialNet paper \cite{Quan2024SpatialNet} augments the training data using RIRs simulated with the open-source toolbox gpuRIR~\cite{DiazGuerra2021gpuRIR}. This toolbox implements GPU-accelerated image-source simulations, and combines them with diffuse late reverberation tails~\cite{DiazGuerra2021gpuRIR}. Following the simulation setup of the original SpatialNet paper, we transition between ISM and diffuse reverberation after the first \SI{15}{\decibel} of energy decay. This ensures a faithful simulation of early reflections, while efficiently modelling the high reflection density characteristic of late reverberation.

For this study, we simulate two ISM-based datasets. The first dataset, ISM-U, randomly samples target reverberation times and room dimensions from predefined intervals. These intervals are chosen to match the minimum and maximum values of the hybrid dataset (see Sec.~\ref{subsubsec:hybrid}) to ensure comparability: $x \in [\SI{3}{\meter}, \SI{33}{\meter}]$, $y \in [\SI{3}{\meter}, \SI{26}{\meter}]$, $z \in [\SI{2.5}, \SI{4.7}{\meter}]$, $T_{20} \in [\SI{0.2}{\second}, \SI{1.6}{\second}]$, where $x$, $y$, and $z$ are the respective room dimensions. Randomly sampling reverberation times and room dimensions is common practice and widely used in state-of-the-art speech enhancement studies. This setup represents an uninformed simulation scenario in which no prior knowledge about typical materials or room volumes is incorporated.

The second dataset, ISM-M, incorporates additional information to obtain more realistic material and room dimension choices. Specifically, we match the setup of the hybrid dataset as closely as possible. To this end, we compute bounding boxes for each room in the hybrid dataset, and use the resulting dimensions in gpuRIR. We also replicate the exact source and receiver positions from the hybrid dataset and set the $T_{20}$ values from the hybrid simulation as the target reverberation times for the gpuRIR simulations. This way, the ISM-M dataset closely matches the hybrid dataset, with the only remaining differences being the presence of scattering objects in the rooms and the simulation and source–receiver modelling paradigm.

For both ISM datasets, the Eigenmike array is modelled as an open microphone array. This approach is common practice when working with gpuRIR. The direct-path speech target signals for training the network are obtained through anechoic simulations for both ISM datasets.

\subsubsection{Hybrid: High-fidelity dataset}
\label{subsubsec:hybrid}
The hybrid dataset was generated entirely using the Treble SDK simulation tool, with all geometry models and boundary-condition materials taken from the included libraries. The dataset consists of three subsets representing different room types and size ranges. The first subset contains 133 living-room geometries with volumes ranging from 40 to \SI{180}{\meter^3}. The second subset contains 103 classroom geometries with volumes ranging from 90 to \SI{400}{\meter^3}. The third subset contains 88 restaurant geometries with volumes ranging from 300 to \SI{1600}{\meter^3}. All geometries are furnished and vary in shape and complexity.

All materials are frequency-dependent and characterised by complex surface impedances. The materials applied to the surfaces in the room models are curated to represent realistic scenarios. For example, windows use glass materials, furniture uses wood or plastic materials, and walls and ceilings use gypsum or concrete materials.

Each room contains four sources placed randomly within the space. Most sources are directive and represent either speech sources or loudspeakers, with their orientations randomised. In larger rooms, one speech source is replaced by an omnidirectional source to increase variability. The number of receivers ranges from 20 to 30 per room, distributed randomly and positioned at least \SI{1}{\meter} from any source and \SI{0.5}{\meter} from any surface. During training, SpatialNet uses scenes of up to 3 overlapping speakers from the same environment. We randomly chose them from the dataset while ensuring that they are all in the same room, resulting in 4801 scenes in total. The distribution of reverberation times is shown in Figure~\ref{subfig:t20_train}. 

The crossover frequency, i.e., the upper frequency limit of the wave solver, is set between 1 and \SI{2}{\kilo\Hz} depending on the room size. The remaining part of the audible spectrum up to \SI{12}{\kilo\Hz}, is simulated using the GA solver of the Treble SDK, which combines the ISM with a ray-radiosity approach. For this dataset, the maximum ISM order is three.

Finally, the effect of the Eigenmike array is incorporated by simulating a full-wave free-field device-related transfer function (DRTF) up to \SI{12}{\kilo\Hz} with the Treble SDK. Using the 16th-order Ambisonics RIRs from the hybrid simulations, the Eigenmike DRTF can be rendered in post-processing. This approach accounts for the full scattering effects of the Eigenmike sphere geometry. From the simulated Eigenmike response, we extract six channels evenly distributed around the sphere and use them for training, as outlined in Section \ref{subsec:network_config}. We obtain direct-path speech target signals by windowing the RIR after the theoretical propagation delay including a \SI{10}{\milli\second} safety margin.



\subsection{Evaluation dataset}
We evaluate all trained models on a dataset of measurements to avoid bias towards either simulation paradigm and to assess performance under real-world conditions. To this end, we introduce LibriCSS-EM6, a dataset inspired by the structure of LibriCSS \cite{Chen2020LibriCSS}, but based on the six-channel Eigenmike subset described in Section \ref{subsec:network_config}. Analogous to LibriCSS, the proposed EM6 variant comprises approximately \num{5000} utterances across six overlap conditions: 0S, 0L, OV10, OV20, OV30, and OV40. The conditions 0S and 0L denote \SI{0}{\percent} speaker overlap with short and long inter-speaker silences, respectively, whereas OV10–OV40 correspond to \SIrange{10}{40}{\percent} speaker overlap. Reverberant speech is generated by convolving dry LibriSpeech utterances from the clean test set \cite{Panayotov2015Librispeech} with Eigenmike RIRs. The LibriCSS metadata provide the necessary time stamps and utterance identifiers to reconstruct the continuous streams, and segmentation follows the original LibriCSS procedure.

The measured Eigenmike RIRs were randomly drawn from two publicly available datasets: Motus~\cite{Goetz2021MotusDatasetPaper} and Arni6DoF~\cite{McKenzie2021Arni6DoF}. The Motus dataset comprises Eigenmike RIRs recorded in a room with various furniture configurations, making it suitable for assessing model performance under different reverberation scenarios and with varying scattering objects. Arni6DoF, recorded in a variable-acoustics room, further extends the evaluation set with additional reverberation conditions. Figure~\ref{subfig:t20_eval} illustrates the resulting band-averaged $T_{20}$ distribution for LibriCSS-EM6, demonstrating broad coverage of reverberation times representative of medium-sized rooms.

Following the LibriCSS structure, the EM6 subset comprises 60 sessions (10 sessions $\times$ 6 overlap conditions). Each session uses random RIRs from either the Motus or Arni6DoF dataset, with equal contributions from both and a balanced assignment across overlap conditions. Arni6DoF comprises five room configurations, each repeated across six sessions, while Motus provides enough configurations to assign a distinct room to every session. Before segmentation, we add multichannel diffuse noise to the speech mixture at a random SNR between 0 to \SI{20}{\decibel}. The noise signals are generated from \mbox{REVERB} Challenge recordings \cite{Kinoshita2013ReverbChallenge} using the method from~\cite{Habets2008GeneratingNonstationarySignalsCoherenceConstraint}.

\subsection{Evaluation setup}
We train one SpatialNet model per training dataset. In all cases, training ran for 30 epochs, after which the validation loss no longer improved substantially. Following the evaluation proposed in the SpatialNet paper \cite{Quan2024SpatialNet}, we average the network weights over the last ten epochs to stabilise ASR performance.

We evaluate the downstream performance of the different \mbox{SpatialNet} variants by transcribing the enhanced speech signals with a Kaldi \cite{Povey2011Kaldi} speech recognition pipeline implemented in pyKaldi2 \cite{Lu2019pyKaldi2}, adopting the same evaluation pipeline that was proposed for the original LibriCSS dataset \cite{Chen2020LibriCSS}. More specifically, acoustic scores are generated by a 3-layer bidirectional long short term memory (BLSTM) \cite{Graves2013BLSTMforASR} acoustic model (512 units per direction) trained on LibriSpeech with cross-entropy and maximum mutual information (MMI) \cite{Bahl1986MMIforSpeechRecognition} sequence training, and decoded with Kaldi using the standard LibriSpeech 4-gram language model.

\begin{figure}
\centering
\includegraphics[width=\columnwidth]{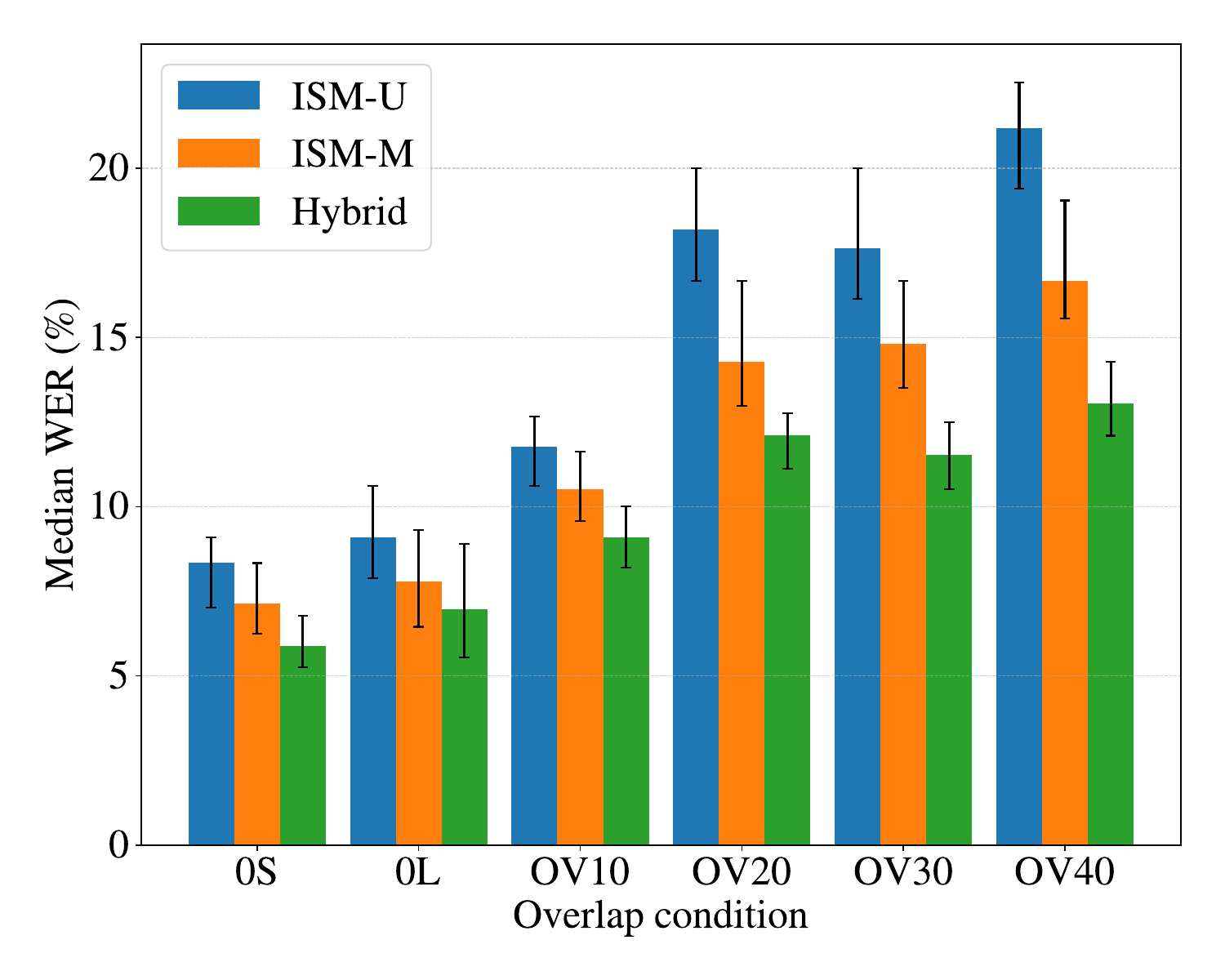}
\caption{Median word error rate (WER) and bootstrapped \SI{95}{\percent} confidence intervals for speech enhanced with different SpatialNet models across several speaker overlap conditions. The models were trained on datasets augmented with room-acoustic simulations of varying fidelity.}
\label{fig:wer_ov}
\end{figure}

\section{Results}
\label{sec:results}
\begin{table}[t]
\centering
\caption{Absolute and relative median WER improvement achieved by the Hybrid training dataset compared to the ISM datasets, with \SI{95}{\percent} bootstrapped confidence intervals. Positive numbers indicate that the Hybrid dataset achieves a better WER than the reference dataset.}
\label{tab:impr-by-condition}
\sisetup{detect-all,table-number-alignment=center,retain-explicit-plus=false}
\setlength{\tabcolsep}{3pt}
\begin{tabular}{l S[table-format=+1.2]@{\,[\hspace{-0.5em}}S[table-format=+2.2]@{,}S[table-format=+1.2]@{]}@{\quad} S[table-format=+2.1]@{\,[\hspace{-0.5em}}S[table-format=+2.1]@{,}S[table-format=+2.1]@{]}}
\toprule
\shortstack{Reference \\ system\\ \quad} & \multicolumn{3}{c}{\shortstack{Absolute\\ median WER\\ improvement}} & \multicolumn{3}{c}{\shortstack{Relative\\ median WER\\ improvement (\%)}} \\
\midrule
\midrule
\addlinespace[0.4em]\multicolumn{7}{l}{\textbf{Overlap condition: 0S}} \\ \addlinespace[0.2em]
ISM-U & 2.17 & 1.26 & 3.03 & 26.7 & 17.2 & 34.2 \\
ISM-M & 1.19 & 0.46 & 2.07 & 16.7 & 6.7 & 25.5 \\
\midrule
\addlinespace[0.4em]\multicolumn{7}{l}{\textbf{Overlap condition: 0L}} \\ \addlinespace[0.2em]
ISM-U & 1.91 & 0.66 & 3.21 & 21.4 & 7.7 & 34.6 \\
ISM-M & 0.72 & -0.55 & 1.88 & 8.7 & -7.7 & 23.5 \\
\midrule
\addlinespace[0.4em]\multicolumn{7}{l}{\textbf{Overlap condition: OV10}} \\ \addlinespace[0.2em]
ISM-U & 2.78 & 1.67 & 3.87 & 23.5 & 15.0 & 31.4 \\
ISM-M & 1.50 & 0.60 & 2.48 & 14.2 & 5.8 & 22.3 \\
\midrule
\addlinespace[0.4em]\multicolumn{7}{l}{\textbf{Overlap condition: OV20}} \\ \addlinespace[0.2em]
ISM-U & 6.15 & 4.57 & 7.50 & 33.3 & 27.3 & 39.0 \\
ISM-M & 2.21 & 1.07 & 4.17 & 15.6 & 8.2 & 25.0 \\
\midrule
\addlinespace[0.4em]\multicolumn{7}{l}{\textbf{Overlap condition: OV30}} \\ \addlinespace[0.2em]
ISM-U & 6.14 & 4.73 & 7.64 & 34.6 & 28.1 & 40.3 \\
ISM-M & 3.20 & 2.25 & 4.67 & 22.2 & 15.6 & 29.0 \\
\midrule
\addlinespace[0.4em]\multicolumn{7}{l}{\textbf{Overlap condition: OV40}} \\ \addlinespace[0.2em]
ISM-U & 8.18 & 6.67 & 9.56 & 38.3 & 33.3 & 43.2 \\
ISM-M & 4.00 & 2.77 & 5.70 & 23.5 & 17.2 & 30.0 \\
\midrule
\midrule
\addlinespace[0.4em]\multicolumn{7}{l}{\textbf{Overall}} \\ \addlinespace[0.2em]
ISM-U & 4.29 & 3.33 & 4.53 & 30.0 & 25.0 & 31.7 \\
ISM-M & 1.93 & 1.43 & 2.50 & 16.3 & 12.9 & 20.0 \\
\bottomrule
\end{tabular}
\end{table}

We enhance and transcribe all utterances of LibriCSS-EM6 with each of the trained SpatialNet models and calculate the corresponding word error rates (WERs). Figure~\ref{fig:wer_ov} shows the median WER of all models under the investigated overlap conditions, with bootstrapped \SI{95}{\percent} confidence intervals. Across all overlap conditions, the model trained with hybrid room-acoustic simulations performs best, whereas the models trained with image-source-based augmentation perform worse. Among the image-source datasets, the variant using informed materials and room dimensions performs considerably better than the one where both parameters were randomly drawn from typical ranges. WER values of the unprocessed noisy+reverberant utterances range from \SI{73}{\percent} (0L condition) to \SI{88}{\percent} (OV40 condition) and are omitted from the figure to highlight differences between training datasets. 

While Figure~\ref{fig:wer_ov} illustrates the overall WER trends across overlap conditions, Table~\ref{tab:impr-by-condition} provides a direct comparison of the models. The table reports the absolute and relative median WER improvements of the model trained with high-fidelity data (Hybrid) over the remaining models. The corresponding bootstrapped \SI{95}{\percent} confidence intervals are computed from paired utterance-level differences, accounting for the shared variability across test utterances. The positive values indicate that the model trained with hybrid room-acoustic simulations consistently outperforms the models trained using image-source-based augmentations. All but one confidence interval lie entirely above zero, indicating that the corresponding improvements are statistically significant at the \SI{95}{\percent} confidence level.

\section{Conclusion}
\label{sec:conclusions}
This paper investigated different room-acoustic simulation paradigms for augmenting training data in multichannel speech enhancement and analysed the impact of simulation fidelity on downstream performance. We observe a relative reduction in median word error rate of up to \SI{38}{\percent} when the training dataset is augmented using high-fidelity room-acoustic simulations rather than simplified approaches. These findings demonstrate that performance gains can be achieved without altering the network or training strategy. Increasing the physical accuracy of the training data alone is sufficient.

\clearpage
\section{Generative AI use disclosure}
The authors acknowledge the use of ChatGPT 5.2 (accessed in February 2026) to rephrase some sentences for clarity, polish the manuscript's language, and help to arrange and format tables and figures. All AI-assisted content was reviewed and revised by the authors to ensure accuracy and clarity of meaning.

\bibliographystyle{IEEEtran}
\bibliography{refs26}

\end{document}